\documentclass[twocolumn,preprint]{aa}
\usepackage[varg]{txfonts}

\usepackage{graphicx}
\usepackage{textcomp}
\usepackage{color}
\usepackage{bbold}
\usepackage{breqn}
\usepackage{amsmath}
\usepackage{pstricks} 
\usepackage{bm} 
\usepackage{relsize}
\usepackage{footmisc}

\usepackage[colorlinks=true,urlcolor=blue,citecolor=blue,linkcolor=blue,breaklinks=true]{hyperref}
\usepackage{twoopt}

\def\xlinkspace#1 #2{%
 \ifx\relax#2%
 \xlinkdash#1-\relax
 \else
 \xlinkdash#1 -\relax
 \expandafter\xlinkspace\expandafter#2%
 \fi}

\def\xlinkdash#1-#2{%
 \ifx\relax#2%
 \tmp{#1}%
 \else
 \tmp{#1-}%
 \expandafter\xlinkdash\expandafter#2%
 \fi}

\bibpunct{(}{)}{;}{a}{}{,} 

\makeatletter

 \newcommandtwoopt{\citeads}[3][][]{%
   \nonstopmode
   \href{http://adsabs.harvard.edu/abs/#3}%
        {\def\hyper@linkstart##1##2{}%
         \let\hyper@linkend\@empty\citealp[#1][#2]{#3}}
   \biblink{#3}{\href{http://adsabs.harvard.edu/abs/#3}{ADS}}%
   \errorstopmode}            
   
 \newcommandtwoopt{\citepads}[3][][]{%
   \nonstopmode
   \href{http://adsabs.harvard.edu/abs/#3}%
        {\def\hyper@linkstart##1##2{}%
         \let\hyper@linkend\@empty\citep[#1][#2]{#3}}
   \biblink{#3}{\href{http://adsabs.harvard.edu/abs/#3}{ADS}}
   \errorstopmode}            
   
 \newcommandtwoopt{\citetads}[3][][]{%
   \nonstopmode
   \href{http://adsabs.harvard.edu/abs/#3}
        {\def\hyper@linkstart##1##2{}%
         \let\hyper@linkend\@empty\citet[#1][#2]{#3}}
   \biblink{#3}{\href{http://adsabs.harvard.edu/abs/#3}{ADS}}%
   \errorstopmode}            
   
 \newcommandtwoopt{\citeyearads}[3][][]{%
   \nonstopmode
   \href{http://adsabs.harvard.edu/abs/#3}%
        {\def\hyper@linkstart##1##2{}%
         \let\hyper@linkend\@empty\citeyear[#1][#2]{#3}}
   \biblink{#3}{\href{http://adsabs.harvard.edu/abs/#3}{ADS}}%
   \errorstopmode}            
\makeatother

\makeatletter
\newcommand{\bibnote}[2]{\@namedef{#1note}{#2}}
\newcommand{\biblink}[2]{\@namedef{#1link}{#2}}
\makeatother

\newcommand{\be}{\begin{equation}}
\newcommand{\ee}{\end{equation}}

\newcommand{\bea}{\begin{eqnarray}}
\newcommand{\eea}{\end{eqnarray}}
\newcommand{\Caline}{\ion{Ca}{ii}~854.2~nm}
\newcommand{\Feline}{\ion{Fe}{i}~617.3~nm}

\begin{document}

\title{Improved reconstruction of solar magnetic fields from imaging spectropolarimetry through spatio-temporal regularisation}
\titlerunning{magnetic field determination with spatio-temporal regularisation}
\authorrunning{de la Cruz Rodr\'iguez and Leenaarts}

\author{
  J. de la Cruz Rodr\'{i}guez\inst{1} \and
  J. Leenaarts\inst{1}
}

\offprints{J. de la Cruz Rodr\'iguez \email{jaime@astro.su.se}}

\institute{Institute for Solar Physics, Dept. of Astronomy, Stockholm University, AlbaNova University Centre, SE-106 91 Stockholm, Sweden}

\date{Received; Accepted }

\abstract 
{Determination of solar magnetic fields with a spatial resolution set by the diffraction limit of a telescope is difficult because the time required to measure the Stokes vector with sufficient signal-to-noise is long compared to the solar evolution timescale. This difficulty gets worse with increasing telescope size as the photon flux per diffraction-limited resolution element remains constant but the evolution timescale decreases linearly with the diffraction-limited resolution.}
{We aim to improve magnetic field reconstruction at the diffraction limit without averaging the observations in time or space, and without applying noise filtering. } 
{The magnetic field vector tends to evolve slower than the temperature,  velocity and microturbulence. We exploit this by adding temporal regularisation terms for the magnetic field to the linear least-squares fitting used in the weak-field approximation, as well as to the Levenberg-Marquardt algorithm used in inversions. The other model parameters are allowed to change in time without constraints. We infer the chromospheric magnetic field from \Caline\ observations using the weak field approximation and the photospheric magnetic field from \Feline\ observations, both with and without temporal regularisation.}
{Temporal regularisation reduce the noise in the reconstructed maps of the magnetic field and provides a better coherency in time in both the weak-field approximation and Milne-Eddington inversions.}
{Temporal regularisation markedly improves magnetic field determination from spatially and temporally resolved observations.}

    \maketitle

\section{Introduction} \label{sec:intro}
The degree of polarisation in spectral lines induced by magnetic fields in the solar atmosphere can be higher than 10\% in regions with strong field
\citepads[e.g.][]{1990ApJ...361L..81M,2020ApJ...895..129C},
but more commonly it is much lower:  
linear polarisation induced by the Zeeman and Hanle effect  in lines formed in the chromosphere tends to be smaller than 0.1\%, especially in the quiet Sun where field strengths are low
\citepads[e.g.][]{2018A&A...619A..60J,2021ApJ...918...15C}.
Observations of the full Stokes vector therefore require long integration times and/or spatial averaging over areas larger than the diffraction limit of the telescope in order to reach the signal-to-noise ratio (SNR) required to detect signals with a strength of  <0.1\% of the local continuum intensity
\citepads[e.g.][]{2010ApJ...708.1579C}.

A major goal of modern solar telescopes is to perform imaging spectropolarimetry at the diffraction limit. This is reflected in the calibration accuracy of the instruments. For example,  the CRISP instrument at the Swedish 1-m Solar Telescope 
\citepads{2003SPIE.4853..341S}
can currently reach a polarimetric accuracy of $\sim 10^{-3}$ and the DKIST telescope can be calibrated to an accuracy of $5\times 10^{-4}$ 
\citepads{2020SoPh..295..172R}.

Because of the finite brightness of the Sun, this SNR is not reached instantaneously. For example, it takes 2~s to reach a SNR of $5\times 10^{-4}$ at the diffraction limit in the continuum adjacent to the \Caline\ line at a spectral resolution $\lambda/\Delta\lambda= 80,000$ and a total transmission of the telescope-instrument-camera chain of 5\%. Tuneable-filter-based instruments like CRISP or VTF at DKIST require data acquisition typically at 10 to 20 wavelengths, depending on the width of the line and the desired wavelength coverage. This leads to a total acquisition time of roughly~20 to 40\,s for a diffraction-limited full-Stokes line scan.

This time is large compared to the timescale on which the density, temperature and velocity in the atmosphere change: the sound crossing time for a diffraction-limited pixel of a 1~m telescope at 854 nm is 9~s assuming a sound speed of 7~km\,s$^{-1}$, for a 4-m telescope it is 2~s. The sound speed should be considered as a lower limit, in the chromosphere the Alfv\'en speed is a better indicator of the evolution speed. It can be larger than 100~km\,s$^{-1}$, implying changes of the thermodynamic parameters in less than~0.1~s.

Thus, using diffraction-limited instruments based on tuneable filters to reach the required high SNR inevitably leads to observed Stokes profiles that average different atmospheric states in density, temperature, and velocity. Using integral field spectropolarimeters such as microlensed hyperspectral imagers 
\citepads{2022A&A...668A.149V}
or image slicers
\citepads{2022JAI....1150014D}
mitigate \,--\,but not fully eliminate\,--\,this effect, but at the price of a strongly reduced field-of-view (FOV).

This mismatch of required integration times and evolution timescales poses a severe problem for inferring chromospheric magnetic fields, especially over the large FOVs of Fabry-P{\'e}rot-type instruments. The equations that describe radiative transfer are non-linear. An inversion of a time-averaged line profile does in general not yield an inferred atmosphere that represents the time-average of the underlying atmosphere 
\citepads{1995ApJ...440L..29C}.
Time-averaging of the Stokes vector might even result in an observed average Stokes vector that cannot be reproduced by any single atmosphere model.

The magnetic field in the chromosphere, where the magnetic pressure is larger than the gas pressure, tends to be rather smooth and slow-varying over space.
\citetads{2020A&A...642A.210M}
realised that this property can be exploited to improve inferring magnetic fields from noisy data using the weak-field approximation (WFA). They extended the merit function used in the WFA-fit with a term that penalises differences in magnetic field in neighbouring pixels, and showed that this leads to substantially improved inference of the longitudinal magnetic field based on the \ion{Mg}{i}~517.3\,nm, \ion{Na}{i}~589.6\,nm, and \Caline\ lines.

The magnetic field in the chromosphere generally only evolves over granular turnover timescales of several minutes
\citepads{2017ApJ...834...26K,2023ApJ...954..185F},
but the mass density, temperature, and velocity evolve with a timescale of seconds on spatial scales of a few tens of kilometers. We propose to exploit this temporal smoothness to improve magnetic field inferral from observation taken at or close to the diffraction limit.

The general idea is to perform inversions where, for the same spatial pixel, observations at consecutive timesteps are allowed to have arbitrary variations in the velocity, temperature and other variables that are well-constrained by Stokes $I$ at relatively low SNR. The magnetic field, which is constrained by Stokes $Q$, $U$, and $V$ and requires a high SNR to be constrained, should vary as little as possible while staying consistent with the data, given the noise level.

This concept can be viewed as applying Tikhonov regularisation 
(\citeads{Tikhonov77}; \citeads{doi:10.1137/1021044})
 to the magnetic field in the time domain. In the case of the WFA this can be implemented as a straightforward extension of the method presented by  
\citetads{2020A&A...642A.210M}. 
In the case of inferral methods based on explicit radiative transfer, whether in the Milne-Eddington approximation, in LTE, or in full non-LTE, it can be implemented as an extension of the method described in 
 \citetads{2019A&A...631A.153D}.
 The former method is linear and requires only a single matrix inversion to yield a solution; the latter method is non-linear and can be solved using the Levenberg-Marquardt algorithm.
 
In this paper we present implementations for both cases. As a proof of concept we apply the temporally-regularised WFA to observations of quiet Sun in the \Caline\ line, and apply temporally regularised Milne-Eddington inversions to observations of quiet Sun in the \Feline\ line. We chose to illustrate the non-linear case with the Milne-Eddington approximation applied to a photospheric line because it is easy to implement and fast to invert. The formalism remains identical also for LTE or non-LTE inversions of chromospheric lines.

The observations that we use here are obtained with a Fabry-Perot tuneable-filter instrument. The individual line positions contained in the data are not taken simultaneously, but sequentially. We assume that one line scan comprises a timestep that will be represented by one atmosphere model. As we argued, this is an approximation as the solar evolution timescale is smaller than the scan time, and this can translate into errors in the inferred model parameters \citep[e.g.,][] {2018A&A...614A..73F}. We stress that temporal regularisation can also be used for integral-field spectropolarimeters. They record all wavelengths simultaneously, and the integration time used in a timestep is mainly set by the desired SNR, and this time is significantly shorter than would be needed by an tuneable-filter instrument.

\begin{figure*}[!ht]
\centering
\includegraphics[width=\textwidth]{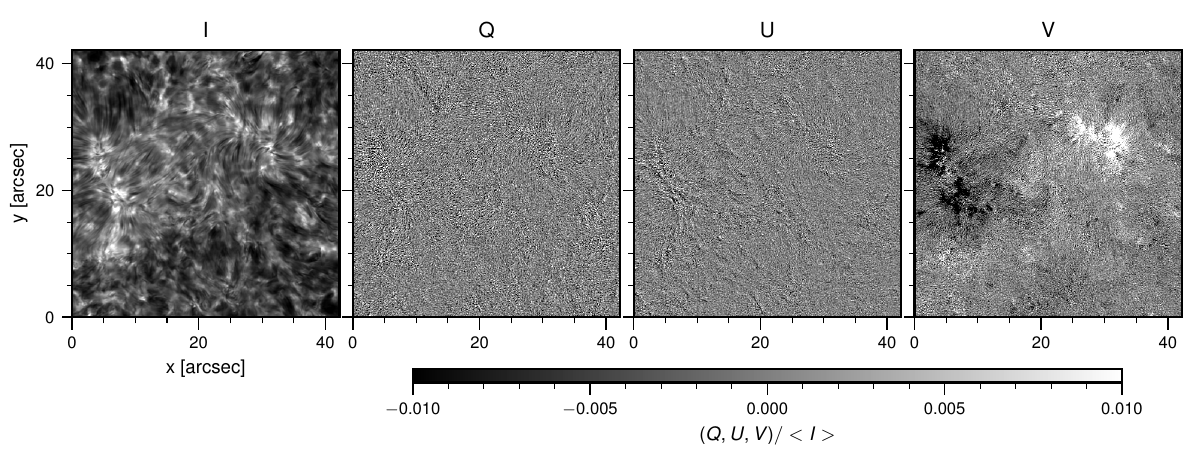}
\includegraphics[width=\textwidth]{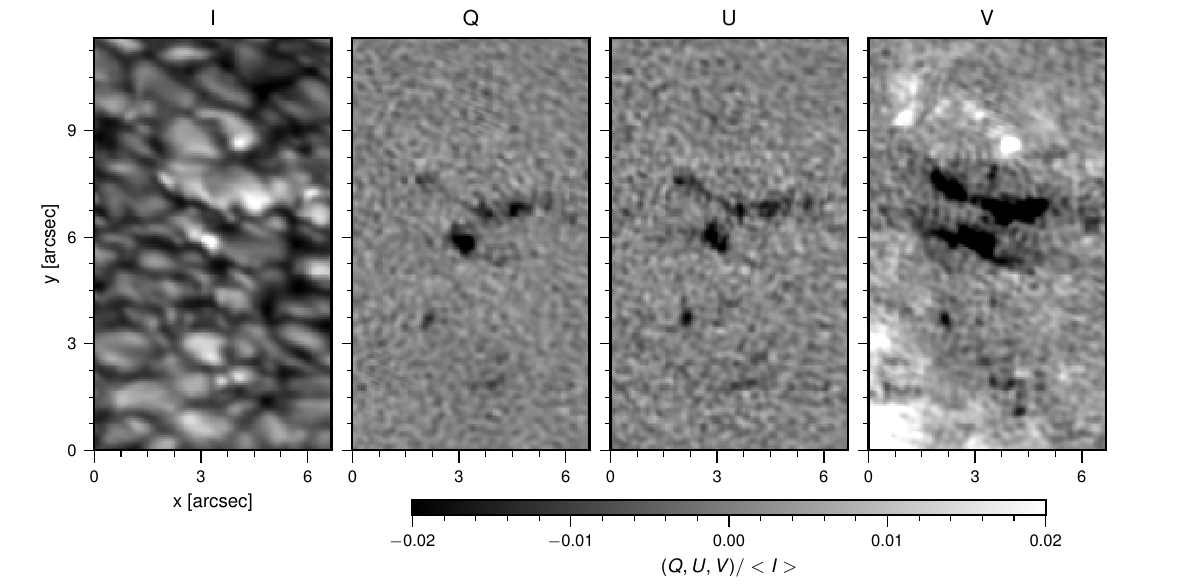}
\caption{Observations used in this study. \emph{Top:} Stokes parameters in the \Caline\ line at $+15$~pm from line center. \emph{Bottom:}  Stokes parameters in the \Feline\ line at $-17.5$ ($I$) and $-7.0$~pm ($Q$, $U$, $V$) from line center. Stokes~$Q$, $U$ and $V$ have been normalised by the spatially-averaged Stokes $I$ intensity at the same wavelength. }
\label{fig:FOV12}
\end{figure*}

\section{Observations}\label{sec:obs}

We  use two datasets, acquired with the CRISP instrument (see 
\citeads{2006A&A...447.1111S}; 
\citeads{2019A&A...626A..55S};
\citeads{2021AJ....161...89D})
at the Swedish 1-m Solar Telescope. The datasets were reduced with the SSTRED pipeline 
(\citeads{2015A&A...573A..40D}; \citeads{2021A&A...653A..68L}), 
and processed with the MOMFBD algorithm to minimise atmospheric distortions 
(\citeads{2002SPIE.4792..146L}; 
\citeads{2005SoPh..228..191V}). 
The polarimetric calibration was individually performed for each pixel as described in \citetads{2008A&A...489..429V}.

The first dataset, recorded on 2020-07-14 starting at 08:40 UT at disk center, consists of a four-line program with data acquired in the \Caline, \Feline, H$\alpha$ and \ion{Mg}{i}~517.3~nm lines. We only used the 854.2~nm data in this study. The line was sampled between $\pm 30$~pm from line center in regular steps of 7.5~pm, with a total integration time per wavelength of $0.91$~s. The overall cadence, including all four lines is $34.2$~s, whereas the total acquisition time for one 854.2~nm scan was approximately 13~s.  The upper row in Fig.~\ref{fig:FOV12} illustrates the four Stokes parameters in the first time-step at $+15$~pm from line center. The target is a quiet-Sun region that harbours two small network patches with opposite polarity. We estimated the noise level based on the outermost wavelength points. The standard deviations of the noise were found to be $\sigma_{Q,U,V} = (3.9,\ 2.6,\ 2.7)\times 10^{-3}$.

The second dataset, recorded on 2021-05-26 starting at 09:48 UT on coordinates $(X,Y) = (698,387)$ $(\mu=0.54)$, consists of a patch of  quiet Sun close to active region AR12824. We recorded data in the \Caline\ and \Feline\ lines, but we only used the 617.3~nm data in this study. The \Feline\ line was sampled at $\Delta\lambda=[-17.5,-10.5,-7,-3.5,0,3.5,7,10.5,14,17.5]$~pm from line center, with a total integration time per wavelength of $0.28$~s. The selected FOV is illustrated in the bottom row of Fig.~\ref{fig:FOV12} in all four Stokes parameters at $-7$~pm from line center. The background noise as estimated from the continuum point has standard deviations $\sigma_{Q,U,V} = (1.8,\ 2.3,\ 2.5)\times 10^{-3}$.

\section{Method}\label{sec:met}

\subsection{The linear case: an application to the weak-field approximation}\label{sec:lin}

\begin{figure}
\centering
\includegraphics[width=0.95\columnwidth]{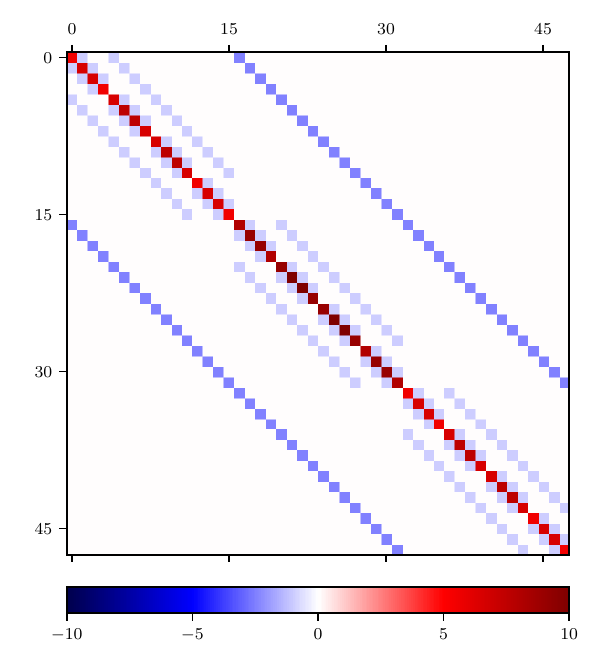}
\caption{Regularisation matrix for the linear case. Each row corresponds to the regularisation function for one pixel in space and time. For displaying purposes, we have assumed an observation with dimensions $n_t=3, n_y=4, n_x=4$ and regularisation weights $\alpha=1$, $\beta=1$ and $\gamma=2.5$. The outermost bands (darker blue) correspond to the temporal regularisation terms whereas the inner light-blue terms originate from the spatial regularisation.}
\label{fig:linmat}
\end{figure}

\begin{figure*}[!ht]
\centering
\includegraphics[width=\textwidth]{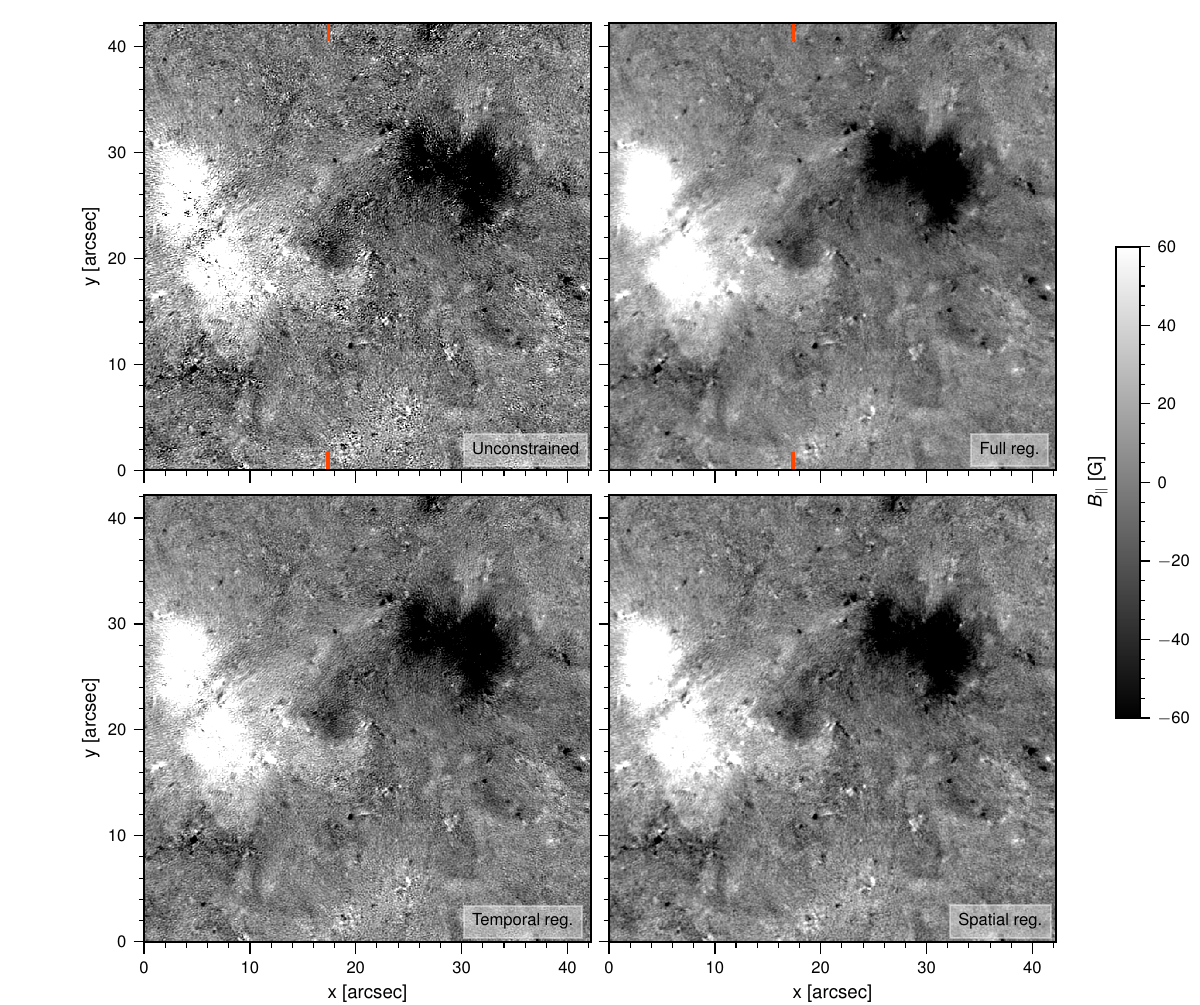}
\caption{Reconstructions of the line-of-sight component of the magnetic field ($B_\parallel$), inferred by applying the weak-field approximation to the \Caline\ data. \emph{Upper-left panel:} reconstruction performed with the unconstrained algorithm.  \emph{Lower-left panel:} reconstruction performed with a temporally-regularised algorithm. \emph{Lower-right panel:} reconstruction performed with a spatially-regularised algorithm. \emph{Upper-right panel:} reconstruction performed with full regularisation in space and time. A movie with the time series is available \href{https://dubshen.astro.su.se/~jaime/movies_paper/fig_WFA.mp4}{online}. The red tickmarks indicate the location of the cut shown in Fig.~\ref{fig:slice}.}
\label{fig:wfa}
\end{figure*}

When the dependence of the observable with the model parameters is linear, spatial and temporal regularisation can be trivially imposed. The weak-field approximation 
(\citeads{1973SoPh...31..299L}; \citeads{1989ApJ...343..920J}) 
allows writing analytical expressions for the $Q$, $U$, and $V$ profiles that are linear in $B_\perp^2$ and $B_\parallel$.

We extend the implementation of \citetads{2020A&A...642A.210M} by also adding regularisation in the temporal dimension. In order to derive general expressions for the linear case, let us assume that we can express our synthetic prediction $s_i$ of the $i-th$ data point in a pixel as $s_i = c_i P$, where $c_i$ is a quantity that does not depend on the model parameters (but that can change for each data point) and $P$ is a model parameter. For a given pixel at coordinates $t,y,x$, we can write the regularised merit function $\chi^2$ as:
\begin{dmath}
    \chi^2 = \frac{1}{N_{\mathrm{data}}}\sum_{i=1}^{N_{\mathrm{data}}}\left(\frac{o_{t,y,x}^i - c_iP_{t,y,x}}{\sigma_i}\right)^2  + \alpha \left[ (P_{t,y,x} - P_{t,y,x-1})^2 + (P_{t,y,x} - P_{t,y,x+1})^2 + (P_{t,y,x} - P_{t,y-1,x})^2+ (P_{t,y,x} - P_{t,y+1,x})^2\right ] + \beta \left (P_{t,y,x} - P_0\right)^2  + \gamma \left [(P_{t,y,x} - P_{t-1,y,x})^2 + (P_{t,y,x} - P_{t+1,y,x})^2 \right ],\label{eq:chi2lin}
\end{dmath}
where $o_i$ is the $i-th$ observed data point, $\alpha$ is a spatial regularisation weight, $\beta$ is a low-norm regularisation weight that penalises deviations from a constant value $P_0$ and  $\gamma$ is a temporal regularisation weight. Taking the derivative of Eq.~\ref{eq:chi2lin} with respect to all $P_{t,y,x}$, and equating it to zero to find the minimum, while absorbing the scaling factors $N_{\mathrm{data}}$ 
and $\sigma_i$ into $o_i$ and $c_i$, we can derive a sparse linear system of equations ($\mathbf{A}\boldsymbol{P} = \boldsymbol{B}$), where the different temporal and spatial locations are coupled by the non-diagonal regularisation terms:
\begin{dmath}
  \left [ \sum_i c_i^2 + 4\alpha + \beta \right]P_{t,y,x} - \alpha\left(P_{t,y-1,x} + P_{t,y+1,x} + P_{t,y,x-1} + P_{t,y,x+1} \right) - \gamma \left ( P_{t-1,y,x} + P_{t+1,y,x} \right) = \sum_i c_io_{t,y,x}^i + \beta P_0. \label{eq:lincoupled}
\end{dmath}
Equation~\ref{eq:lincoupled} couples the parameters of the different pixels ($P_{t,y,x}$) through the regularisation weights ($\alpha$, $\beta$, $\gamma$) on the left-hand side.
Obtaining a solution requires solving a global problem for all pixels simultaneously.
If the regularisation weights are zero, the $\mathbf{A}$  matrix in Eq.~\ref{eq:lincoupled}  is diagonal and we recover the unconstrained algorithm. Figure~\ref{fig:linmat} shows the structure of $\mathbf{A}$ for a small FOV of $4\times4$ pixels and $3$ time steps. Compared to Fig.~1 of \citetads{2020A&A...642A.210M}, this matrix has two extra off-diagonal bands originating from the temporal regularisation terms.

If we replace $P_{t,y,x}$ with $B_{\parallel}^{t,y,x}$ and $\sum_i c_i$ with $\sum_i C \partial I_i / \partial \lambda_i$ we directly obtain the expression for the fully regularised WFA, with $C=4.67\times 10^{-12}\lambda_0^2 g_{\mathrm{eff}}$, $\lambda_0$ the central wavelength of the line in nm and $g_{\mathrm{eff}}$ the effective Land\'e factor of the transition.

\begin{figure}
\centering
\includegraphics[width=\columnwidth]{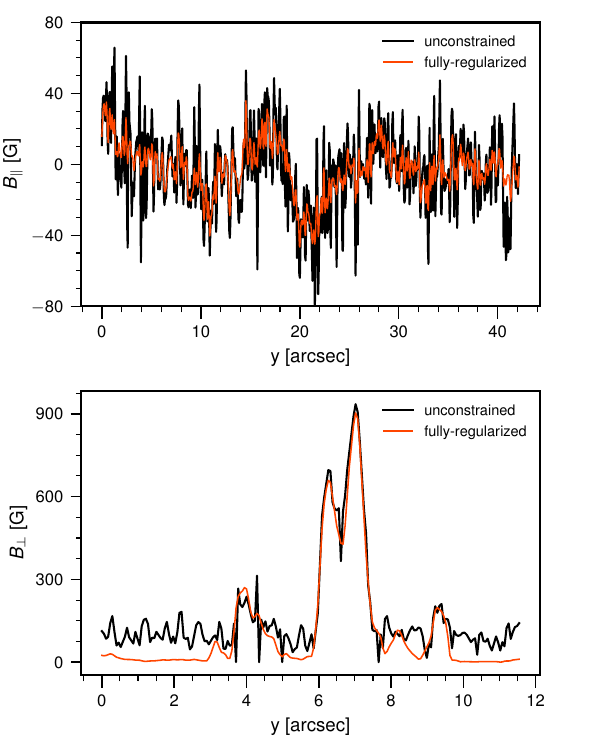}
\caption{Vertical cuts of the reconstructed magnetic field. The upper panel shows a reconstruction of $B_\parallel$ with the weak-field approximation from the 854.2 nm dataset along a cut indicated with red markers in Fig~\ref{fig:wfa}. The black curve shows the unconstrained reconstruction, the red curve the reconstruction with both spatial and temporal regularisation. A similar plot is shown in the lower panel for the reconstruction of $B_\perp$ from the Milne-Eddington inversion of the 617.3~nm dataset presented in Fig.~\ref{fig:ME}.}
\label{fig:slice}
\end{figure}

Figure~\ref{fig:wfa} illustrates a comparison of four reconstructions of $B_\parallel$ from the \Caline\ dataset using the unconstrained WFA algorithm, the spatially regularised method of 
\citetads{2020A&A...642A.210M} , 
and two variants of our method that include temporal regularisation alone as well as both temporal and spatial regularisation. The reconstruction of the unconstrained algorithm has a very noisy background, and besides the magnetic field in the strong network patches,  small-scale loop-like features can be discerned in the FOV. The inclusion of spatial regularisation greatly decreases the noise in the background, making the small-scale loop-like features much more visible compared to the unconstrained case. Temporal regularisation alone also decreases the noise, perhaps yielding a slightly sharper model than the spatially-regularised one, while also decreasing the temporal fluctuations of the background noise along the time series. The combined action of temporal and spatial regularisation further decreases the noise compared to the previous cases and yields a model with the highest SNR. The standard deviation of the noise in the magnetic field, as measured in the lower-right corner of the image, is reduced from $18$~G in the unconstrained case to $9$~G in the fully-regularized case. The improvement induced by spatial regularisation is particularly obvious in the movie showing the entire time-series. The upper panel in Fig.~\ref{fig:slice} further illustrates the reduction of the noise in the regularised reconstructions along a slice through the FOV.

\section{The non-linear case: application to Milne-Eddington inversions}\label{sec:milne}
The Levenberg-Marquardt algorithm (LM, \citeads{10.2307/43633451}; \citeads{marquardt1963}) is one of the most efficient methods to reconstruct the parameters of non-linear models from observations. Regularised LM implementations are used in different fields of astrophysics in order to  constrain the parameters of the model under consideration (e.g., \citeads{2002A&A...381..736P}; \citeads{2019A&A...623A..74D}). 
The regularisation is represented by a set of $N_{\mathrm{pen}}$ penalty functions $r_n(\boldsymbol{p})$ that are squared in order to have $\ell-2$ norm regularisation. Following the notation of \citetads{2019A&A...623A..74D}, the merit function $\chi^2$ can generally be expressed as a function of the model parameter vector $\boldsymbol{p}$ and the data points $\boldsymbol{x}$:
\begin{equation}
  \chi^2(\boldsymbol{p},\boldsymbol{x}) = \frac{1}{N_{\mathrm{data}}}\sum_{k=1}^{N_{\mathrm{data}}} \left[\frac{o_k-s(\boldsymbol{p},x_k)}{\sigma_k} \right]^2 + \sum_{n=1}^{N_{\mathrm{pen}}}\alpha_nr_n(\boldsymbol{p})^2,\label{eq:chinon}
\end{equation}
where the weights $\alpha_n$ regulate the influence of the regularisation terms in the merit function. We note that, unlike in the linear case, the penalty functions are not \emph{independently} defined for each pixel (see below).

The model corrections predicted by the regularised Levenberg-Marquardt algorithm can be derived by linearising Eq.~\ref{eq:chinon} and taking the derivative with respect to the model parameters. The correction to the model parameters ($\Delta\boldsymbol{p}$), in vector form, is given by a linear system of equations:
\begin{equation}
(\mathbf{J}\cdot\mathbf{J}^T + \mathbf{L}\cdot\mathbf{L}^T)\Delta\boldsymbol{p} = \mathbf{J}\cdot(\boldsymbol{o}-\boldsymbol{s})-\mathbf{L}\cdot\boldsymbol{r},\label{eq:corr}
\end{equation}
where $\mathbf{J}$ is the Jacobian matrix of the model and $\mathbf{L}$ is the Jacobian matrix of the penalty functions. In this expression, all constants and normalising factors are implicitly contained in the corresponding vectors. For any given parameter $p_{t,y,x}$, the regularisation functions are defined as:
\begin{dmath}
\sum \boldsymbol{r}_{t,y,x}^2 =  \alpha_t(p_{t,y,x} - p_{t-1,y,x})^2 + \alpha_s(p_{t,y,x} - p_{t,y-1,x})^2 + \alpha_s(p_{t,y,x} - p_{t,y,x-1})^2 + \alpha_l(p_{t,y,x}-p_0)^2,
\end{dmath}
where we have included spatial, temporal and low-norm regularisation terms.

\begin{figure}
\centering
\includegraphics[width=0.94\columnwidth]{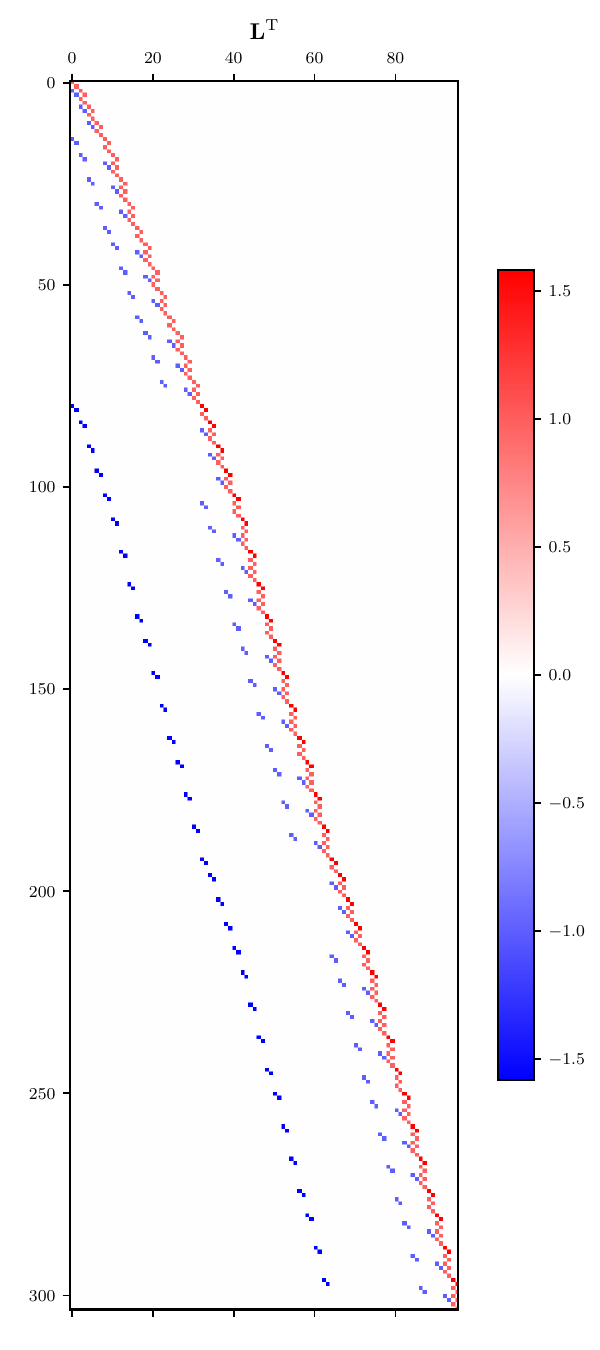}
\caption{Transpose of the Jacobian matrix of the penalty functions for the non-linear case with $n_t=3$, $n_y=4$, $n_x=4$, $n_{\mathrm{par}}=2$ and regularisation weights $\alpha=1$, $\beta=1$ and $\gamma=2.5$. Each row corresponds to one penalty function and each column indicates which free parameters are affected.}
\label{fig:gam}
\end{figure}

\begin{figure}
\centering
\includegraphics[width=0.95\columnwidth]{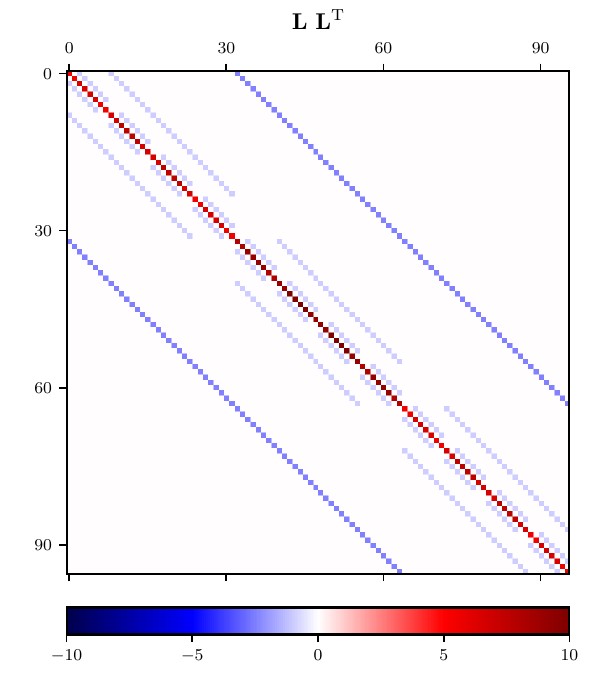}
\caption{Hessian approximation of the regularisation functions ($\mathbf{L}^{T}\mathbf{L}$). This matrix was computed using $\mathbf{L}$ from Fig.~\ref{fig:gam}.}
\label{fig:LL}
\end{figure}

If the penalty functions have a linear dependence with the model parameters, like the ones used in this study, we can express them as $\boldsymbol{r}=\mathbf{L}\boldsymbol{p}+\boldsymbol{c}$, where the coupling is contained in the Jacobian matrix $\mathbf{L}$. Figure~\ref{fig:gam} illustrates the structure of $\mathbf{L}$ for a problem with $n_t=3$, $n_y=4$, $n_x=4$ and $n_{\mathrm{par}} = 2$. The maximum number of penalty functions should be $n_{\mathrm{pen}} < 4n_xn_yn_t$ as some functions are not defined at the edges of the problem. Figure~\ref{fig:LL} illustrates the approximate Hessian matrix of the regularisation functions $\mathbf{L}\cdot\mathbf{L}^{T}$. The size of the approximate Hessian matrix is set by the total number of free parameters of the problem, which is also the size of one row of the $\mathbf{L}^{T}$ matrix ($4\times 4\times 3\times 2=96$). In the latter, the outer dark blue bands originate from the temporal regularisation whereas the inner light blue bands contain the spatial coupling terms. For $n_{\mathrm{par}}=1$, this matrix has identical form to the one shown in Fig.~\ref{fig:linmat} for the linear case. 

In this case we are only including penalty functions that compare a parameter value with its neighbouring values at $t-1$, $y-1$ and $x-1$. However, because the product $\mathbf{L}\cdot\mathbf{L}^T$ is similar to a correlation of each penalty function with all others, the resulting coupling matrix in the left-hand side has identical structure as the one in the linear case where we also included explicit comparisons with the values at $t+1$, $y+1$ and $x+1$.


\begin{figure*}
\centering
\includegraphics[width=0.95\textwidth, trim=0 0.55cm 0 0, clip]{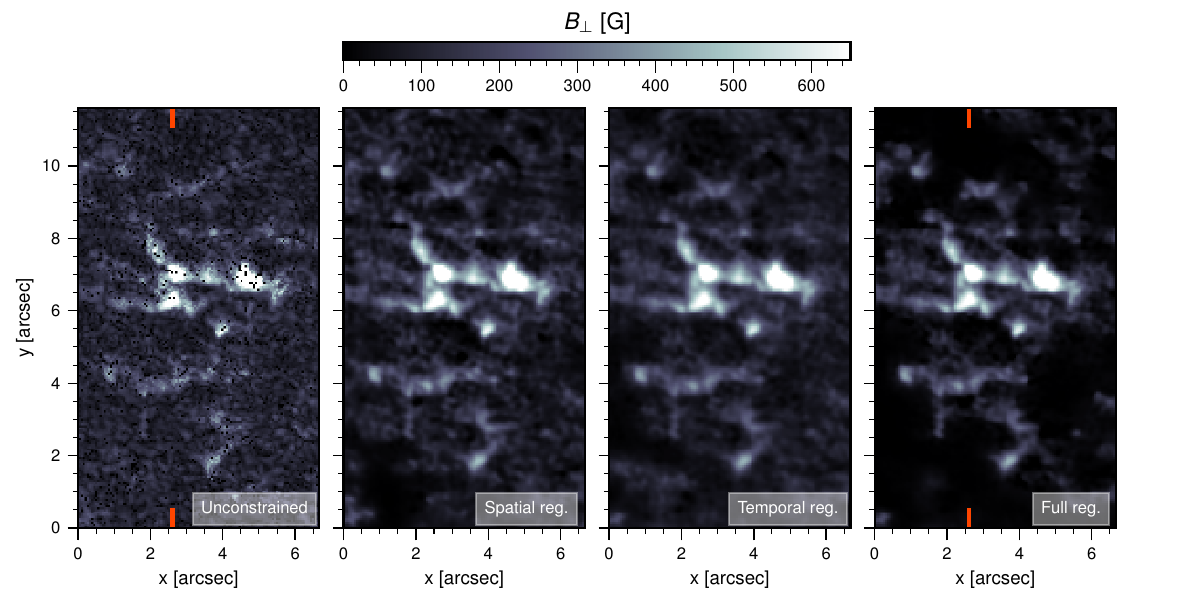}
\caption{Reconstruction of the transverse component of the magnetic field vector from Milne-Eddington inversions of the \Feline. \emph{Left:} Unconstrained reconstruction where each pixel has been inverted six times with five different random initialisations. \emph{Middle-left:} reconstruction with spatial regularisation where all pixels where only inverted once. \emph{Middle-right:} reconstruction with temporal regularisation. \emph{Right:} reconstruction including temporal, spatial and low-norm regularisation terms. A movie with the time-evolution is available \href{https://dubshen.astro.su.se/~jaime/movies_paper/fig_ME_inversion.mp4}{online}.}
\label{fig:ME}
\end{figure*}
In a previous study, \citetads{2019A&A...631A.153D} implemented a multi-resolution inversion code that included spatial regularisation and allowed to deal with the effects of the telescope point-spread-function in Milne-Eddington inversions (PyMilne\footnote{\url{https://github.com/jaimedelacruz/pyMilne}\label{fot:pyMilne}}). We have extended this code to include temporal regularisation terms. With these changes, the code can now invert a time series of maps as a global problem. We inverted a time-series of 14 line scans acquired in the \Feline\ line in a quiet-Sun patch. These observations are not ideal because they also included scans in the \Caline\ line, inducing a time gap between consecutive \ion{Fe}{i} scans. Nevertheless, the cadence is sufficiently high to illustrate the advantages of the method. In the following, we focus on the reconstruction of $B_{\perp}$, which is a more challenging problem than reconstructing $B_{\parallel}$.

We performed reconstructions with the unconstrained algorithm, including spatial regularisation only, temporal regularisation only and a case including low-norm, spatial and temporal regularisation altogether. Regularisation was only imposed in the three components of the magnetic field, leaving all other parameters unconstrained. Therefore, the thermodynamical quantities of the model are allowed to change freely in time and space.

Figure~\ref{fig:ME} compares reconstructions of the time-series of $B_\perp$ from the unconstrained algorithm and the three regularised cases. The improvements of the fully-regularised reconstruction are threefold:
\begin{itemize}
\item so-called inversion noise, caused by pixels where the algorithm did not converge to the global minimum, is absent. This effect is clearest in the patches of strong magnetic field, where randomly-located black pixels are present in the unconstrained reconstruction.  Examples of inversion noise are also visible in the lower panel of Fig.~\ref{fig:slice} at $y\approx 3.7\arcsec$ and $y\approx 4.35\arcsec$.
Inversion noise can be removed from the unconstrained method by initialising the inversion several times with different starting model parameters or by performing a two-cycle inversion with a smoothing step in between (as suggested by \citeads{2018A&A...617A..24M}). However, the noise is naturally addressed in all the regularised cases, where the coupling between pixels in space and/or time removes it entirely.

\item a reduction of the background noise. Such a background \emph{bias}-level can appear in the presence of noise for parameters that are defined positive (see \citeads{2012MNRAS.419..153M}). We find that all forms of regularisation decrease this bias, but the low-norm regularisation term (with $P_0=0$) and the temporal regularisation are particularly effective in this regard. The spatially-regularised time-series shows a stronger frame-to-frame fluctuation in the background level, which seems to be induced by varying atmospheric seeing conditions along the series.
This effect is also visible in the lower panel of Fig.~\ref{fig:slice}, where the red curve is regularly predicting much lower values of $B_\perp$ in regions located between stronger network patches. 

\item the time coherence of weak magnetic structures is enhanced. Structures with signal barely over the noise appear more persistent in time with temporal regularisation. Without this regularisation they appear intermittently, only in those time-steps where the noise happens to be low enough for the inversion to pick up the structure.  

\end{itemize}

However, despite these improvements, the evolution of magnetic fields in the photosphere is anchored to the forces originating from gas pressure gradients (plasma $\beta \gg 1$), and our assumption of having a slowly evolving magnetic field relative to the rest of parameters is not optimal. Compared to the chromospheric case presented in \S\ref{sec:lin}, we had to limit the amount of temporal regularisation to perform these inversions.


\section{Discussion \&\ conclusions}\label{sec:disc}

We present a method to include Tikhonov regularisation in the time dimension for linear and non-linear fits of model parameters. We show that the method is suitable for improving determinations of the magnetic field in the solar chromosphere.

Compared to other techniques that operate on the data to decrease noise, such as temporal and spatial rebinning, Principal Component Analysis filtering (see, e.g., \citeads{2008A&A...486..637M}), Fourier filtering, or neural-network-based noise reduction (\citeads{2019A&A...629A..99D}), regularisation leaves the data untouched and operates only on the model parameters. This is an important advantage because filtering and/or averaging the data likely affects the reconstruction of all parameters, whether that is desired or not. 

Temporal regularisation addresses the different time scales on which the magnetic field and the thermodynamical parameters evolve in the chromosphere as well as the different SNR required to constrain them. Because the thermodynamic parameters evolve relatively fast but require a low SNR, one can use zero (as we do here) or weak regularisation. The magnetic field evolves more slowly but requires high SNR, so it is regularised more strongly. 

Choosing adequate regularisation weights is crucial to obtain results that are not overly smooth. The weights must be chosen in such a way that the final quality of the fit is not affected, but just around that threshold when the quality of the fits starts to be affected by the limitations of the regularisation terms (\citeads{2017A&A...597A..58K}). For observations with low SNR, this might mean that the regularisation is so strong it makes the magnetic field almost constant in time. We emphasise that with properly chosen weights, regularisation does not impose that the magnetic field is constant over time. If the SNR is sufficiently high it still allows  sharp gradients in time. However, in cases of intermediate SNR it will produce smoother solutions if strong gradients are not required to minimise $\chi^2$. Regularisation has an advantage over smoothing the model parameters after an unconstrained inversion, because smoothing washes out variations regardless of the impact that they have on $\chi^2$, but regularisation keeps sharp gradients in locations where they have a significant impact in $\chi^2$.

Most observational programs with Fabry-Perot instruments that observe multiple lines are done by scanning once through each line and repeating this cycle. Long integration times are needed for chromospheric lines in order to reach sufficient SNR to measure polarisation.  As mentioned in the introduction this takes about 20~s, and the observed line profile is not produced by a single atmosphere as the Sun has evolved over this time. 

Temporal regularisation allows for observing strategies that mitigate this. Instead of a single slow scan one can perform multiple fast scans that each have sufficient SNR to constrain the thermodynamic parameters. These profiles now suffer less from solar evolution and therefore lead to improved reconstructions of the thermodynamic parameters. Temporal regularisation of the magnetic field couples the scans together and increases the "effective" SNR sufficiently to determine the field. These fast scans can be interleaved with observations in other lines (which themselves can consist of multiple scans) as needed, as temporal regularisation does not require a constant cadence. If needed, the temporal regularisation weight can be made smaller for scans that are not performed back-to-back.

Another obvious application is data taken with  integral-field spectropolarimeters. Here the assumptions behind temporal regularisation hold much better, because all spatial and spectral datapoints are taken simultaneously, and the cadence is typically high \citep{2023A&A...673A..11R}.

Data taken at the diffraction limit of a telescope have a limited SNR if the exposure time is kept short enough to prevent solar evolution from blurring the images. Spatio-temporal regularisation of the magnetic field is a way of increasing the effective SNR of the data while keeping exposure times short. For data taken with large-aperture telescopes, such as DKIST (\citeads{2020SoPh..295..172R}) and the planned EST (\citeads{2022A&A...666A..21Q}), this technique will be invaluable.

The version of pyMilne that includes temporal regularisation is available in the official repository\footref{fot:pyMilne}. The regularised WFA code used here is also publicly available\footnote{\url{https://github.com/jaimedelacruz/fullReg_wfa}}. 


\begin{acknowledgements}
The authors of this paper are thankful to the referee for his careful evaluation of the manuscript.
The Institute for Solar Physics is supported by a grant for research infrastructures of national importance from the Swedish Research Council (registration number 2021-00169). The Swedish 1-m Solar Telescope is operated on the island of La Palma by the Institute for Solar Physics of Stockholm University in the Spanish Observatorio del Roque de los Muchachos of the Instituto de Astrof\'isica de Canarias.
JL acknowledges financial support from the Swedish Research council (VR, project number 2022-03535).
No animals were harmed in the making of this manuscript.
This project has been funded by the European Union through the European Research Council (ERC) under the Horizon Europe program (MAGHEAT, grant agreement 101088184). Views and opinions expressed are however those of the author(s) only and do not necessarily reflect those of the European Union or the European Research Council. Neither the European Union nor the granting authority can be held responsible for them.
 \end{acknowledgements}

\bibliographystyle{aa}
\bibliography{references}

\end{document}